\documentclass{eas}
\usepackage{graphicx}
\begin{document}

\title{       The first Galaxy scale hunt for the youngest high-mass protostars} 
\author{T. Csengeri}\address{Max Planck Institute for Radioastronomy, \texttt{csengeri@mpifr-bonn.mpg.de}}
\author{S. Bontemps}\address{OASU/LAB, Universit\'e Bordeaux}
\author{F. Wyrowski$^1$}
\author{K. M. Menten$^1$}
\author{S. Leurini$^1$}
\author{J. S. Urquhart$^1$}
\author{F. Motte}\address{Laboratoire AIM Paris Saclay}
\author{F. Schuller}\address{European Southern Observatory}
\author{L. Testi$^4$}
\author{L. Bronfman}\address{Departamento de Astronom\'{i}a, Universidad de Chile}
\author{H. Beuther}\address{Max Planck Institute for Astronomy}
\author{S. Longmore}\address{Astrophysics Research Institute, Liverpool John Moores University }
\author{B. Commercon}\address{Ecole Normale Superieure de Lyon}
\author{Th. Henning$^6$}
\author{A. Palau}\address{Centro de Radioastronom\'ia y Astrof\'isica, Universidad Nacional Aut\'onoma de M\'exico}
\author{J.\,C. Tan}\address{Departments of Astronomy and Physics, University of Florida }
\author{G. Fuller}\address{Jodrell Bank Centre for Astrophysics, University of Manchester}
\author{N. Peretto}\address{School of Physics \& Astronomy, Cardiff University}
\author{A. Duarte-Cabral}\address{School of Physics and Astronomy, University of Exeter}
\author{A. Traficante$^{11}$}

\begin{abstract}
The origin of massive stars is a fundamental open issue in modern astrophysics. Pre-ALMA interferometric studies reveal precursors to early B to late O type stars with collapsing envelopes of 15--20\,M$_{\odot}$ on 1000--3000 AU size-scales. To search for more massive envelopes we selected the most massive nearby young clumps from the ATLASGAL survey to study their protostellar content with ALMA. Our first results using the intermediate scales revealed by the ALMA ACA array providing 3--5'' angular resolution, corresponding to $\sim\,0.05-0.1$\,pc size-scales, reveals a sample of compact objects. These massive, dense cores are on average two-times more massive than previous studies of similar types of objects. 
We expect that once the full survey is completed, it will provide a comprehensive view on 
the origin of the most massive stars.
\end{abstract}
\maketitle
\section{Introduction}
The origin of stellar masses, in particular of massive stars, is one of the most fundamental open issues of modern astrophysics. In spite of recent progress in our understanding of 
high-mass star-formation, a coherent picture is still lacking of the earliest evolutionary stages, i.e.\,, the onset of collapse and the initial fragmentation of massive, dense cores (MDCs) (e.g.\,Tan {\em et al.\/} \cite{Tan2014}).
Potential precursors of early B to possibly late O stars with $M_{\rm env}\sim10-20$\,M$_{\odot}$
have been frequently revealed (e.g.\,Bontemps {\em et al.\/} \cite{Bontemps2010}, Zhang {\em et al.\/} \cite{Zhang2009}, Rathborne {\em et al.\/} \cite{Rathborne2011}, Wang {\em et al.\/} \cite{Wang2011}, Longmore {\em et al.\/} \cite{Longmore2011}, Palau {\em et al.\/} \cite{Palau2013}, Beuther {\em et al.\/} \cite{Beuther2013}). Their typical sizes of 1000--3000\,AU and low bolometric luminosities make them excellent analogs for the low-mass Class 0 stage (Duarte-Cabral {\em et al.\/} \cite{Duarte-Cabral2013}). 

The larger effective Jeans masses of these protostars can be explained by a combination of turbulence and magnetic fields at small scale (McKee \& Tan \/\cite{MT2003}), or by collapse from larger scales at which Jeans masses are larger due to lower average densities (e.g.\,Hennebelle \& Chabrier \/\cite{HC2008}). Recent numerical models reproduce up to 10\,M$_{\odot}$
 fragments from a 100M$_{\odot}$
 collapsing core (Commer\c con {\em et al.\/} \cite{Commercon2011}), further suggesting that the combined effect of magnetic fields and radiative feedback determines the early fragmentation of massive cores (see also Krumholz {\em et al.\/} \cite{Krumholz2012}). To test formation scenarios the precursors of the more massive objects need to be revealed, which becomes now feasible by performing statistical studies using the capabilities of the Atacama Large Millimeter/submillimeter Array (ALMA).

\section{A sample of the brightest ATLASGAL sources}
The ATLASGAL survey (Schuller {\em et al.\/} \cite{Schuller2009}, Csengeri {\em et al.\/} \cite{Csengeri2014}) is the most sensitive and extensive ground-based unbiased survey of the inner Galaxy at sub-millimeter wavelengths, and provides an unprecedented view on all stages of massive star formation. Over 10 000 compact sources have been identified (Csengeri {\em et al.\/} \cite{Csengeri2014}), and we have made substantial progress in characterising various evolutionary stages of massive clumps by using ancillary radio and mid-infrared data (e.g.\,Urquhart {\em et al.\/} \cite{Urquhart2014}), and assigned distances to a large number of sources (Wienen {\em et al.\/} \cite{Wienen2015}). Given its higher angular resolution (19\rlap{.}{''}2) compared to \emph{Herschel} at submillimeter wavelengths, the ATLASGAL survey is better suited to disentangle the more evolved from the pristine cold sources.

Selected from ATLASGAL, we identified a complete sample of 45 objects which are massive ($>$650\,M$_{\odot}$) and dense, with surface density exceeding the theoretical 1\,g\,cm$^{-2}$ threshold (Krumholz {\em et al.\/} (\cite{Krumholz2008}) for forming high-mass stars. In addition they lack bright mid-infrared embedded objects, suggesting that they are in their earliest evolutionary phase. 
Our selection of these  mid-infrared quiet massive clumps is complete within 4.5 kpc, and to date represents the best potential sites to host the next generation of the most massive stars currently forming in our Galaxy. 

\section{ALMA observations of precursors of the youngest massive protoclusters}

We used ALMA to perform the first systematic survey at high angular-resolution to look for high-mass protostars in this sample of massive clumps. As a first step we present here the results of the Atacama Compact Array (ACA) observations, which provides a 3--5'' angular resolution corresponding to $\sim$0.05--0.1 pc physical scales (Fig.\,1). This size-scale shows the fragmentation from clump to core scales, and reveals a sample of MDCs.

The first result of this survey is to confirm the presence of compact embedded sources towards our selection of mid-infrared quiet ATLASGAL clumps. Only one source out of the entire survey is found to be resolved out lacking embedded compact objects, while the rest all shows at least one centrally concentrated fragment. 

Interestingly, the sample shows limited fragmentation, with only 2--3 MDCs revealed per clump, which is similar to earlier findings by Motte {\em et al.\/} (\cite{Motte2007}) and Bontemps {\em et al.\/} (\cite{Bontemps2010}) in the systematic study of the Cygnus-X region. A first mass estimate shows, however that they are on average twice as massive as the MDCs in Cygnus-X. Among them we also find several candidates to be the most massive MDCs known to date (Csengeri {\em et al.\/} \emph{in prep}).

\begin{figure}
\centering
  \includegraphics[width=0.95\linewidth,clip]{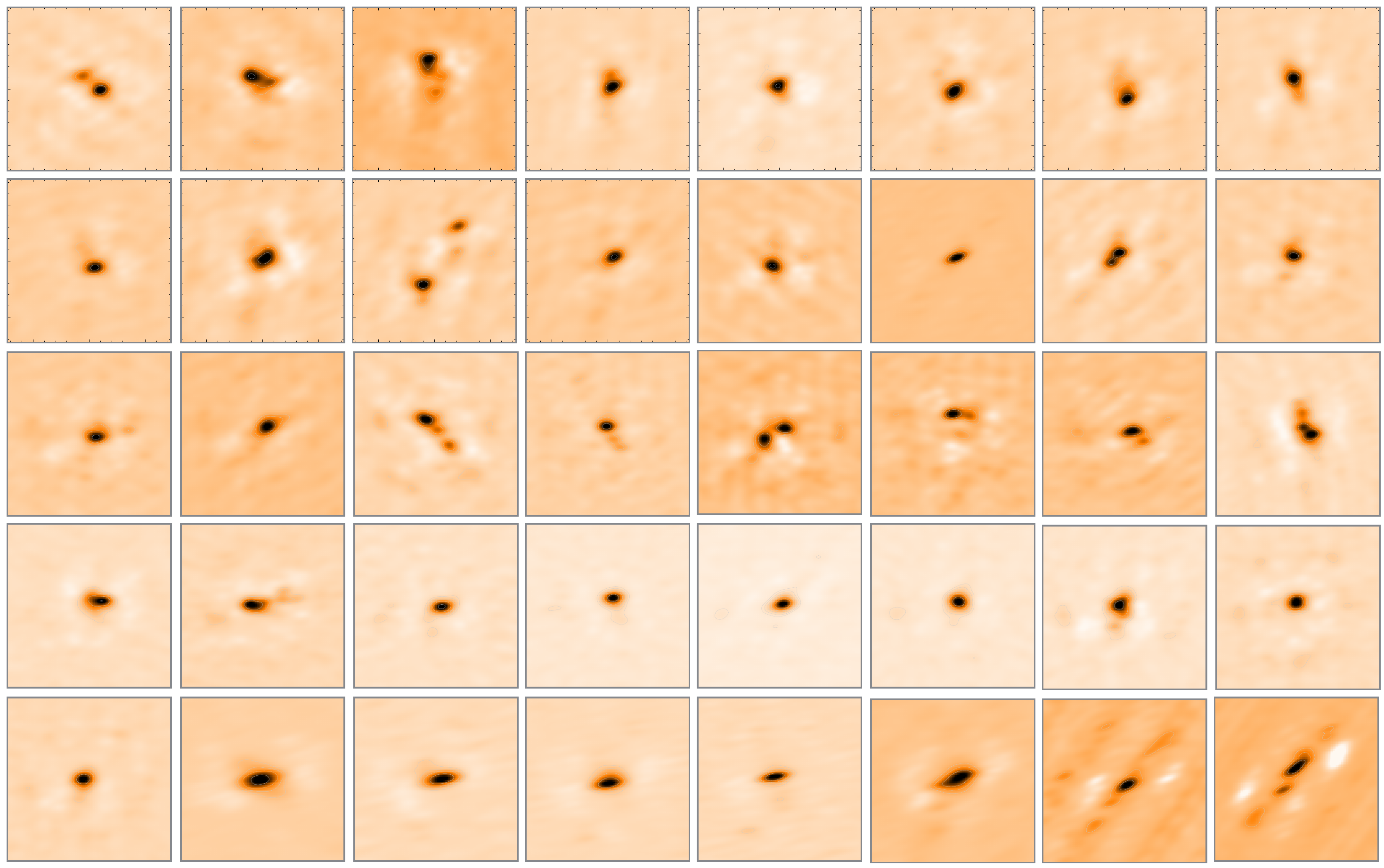}
 \caption{Line-free continuum observations at 870\,$\mu$m of 40 of the targeted sources with 
 ALMA ACA at 3--5'' angular resolution.}
 \label{fig:example}
 \end{figure}

\section{Conclusions and Outlook}

A first analysis of fragmentation on intermediate scales reveals a large number of MDCs within the sample of mid-infrared quiet ATLASGAL sources targeted with ALMA. They are extreme in terms of their mass and surface density, potentially hosting the precursors of the most massive stars in our Galaxy. Complementary molecular line data will be used to further investigate the nature of their embedded sources. The full capacities of ALMA requested for this survey will provide 0.6'' angular resolution, corresponding to individual protostars. This dataset will ultimately provide a comprehensive view of the earliest phase of the formation of high-mass protostars.


\end{document}